\documentclass[sigconf,anonymous=false]{acmart}

\usepackage{booktabs} 
\usepackage{enumitem}
\usepackage{amsmath}
\usepackage{amsfonts}
\usepackage[ruled,vlined]{algorithm2e}
\usepackage{algpseudocode}
\usepackage{bm}
\usepackage{subfigure}
\usepackage{multirow}
\usepackage{hyperref}
\usepackage{fancyhdr}

\AtBeginDocument{%
  \providecommand\BibTeX{{%
    \normalfont B\kern-0.5em{\scshape i\kern-0.25em b}\kern-0.8em\TeX}}}

\copyrightyear{2023}
\acmYear{2023}
\setcopyright{acmlicensed}\acmConference[SIGIR '23]{Proceedings of the 46th International ACM SIGIR Conference on Research and Development in Information Retrieval}{July 23--27, 2023}{Taipei, Taiwan}
\acmBooktitle{Proceedings of the 46th International ACM SIGIR Conference on Research and Development in Information Retrieval (SIGIR '23), July 23--27, 2023, Taipei, Taiwan}
\acmPrice{15.00}
\acmDOI{10.1145/3539618.3591999}
\acmISBN{978-1-4503-9408-6/23/07}

\begin{document}


\title{HyperFormer: Learning Expressive Sparse Feature Representations via Hypergraph Transformer}




\author{Kaize Ding*$^\dagger$,~~~Albert Jiongqian Liang$^\ddagger$,~~~Bryan Perrozi$^\ddagger$,~~~Ting Chen$^\ddagger$,~~~Ruoxi Wang$^\ddagger$,~~~Lichan Hong$^\ddagger$,~~~Ed H. Chi$^\ddagger$,~~~Huan Liu$^\dagger$,~~~Derek Zhiyuan Cheng$^\ddagger$}\thanks{*Work was done as an intern at Google.}

\affiliation{
 \institution{$^\dagger$Arizona State University, \{kaize.ding, huan.liu\}@asu.edu \\ $^\ddagger$Google, \{jiongqian, hubris, iamtingchen, ruoxi, lichan, edchi, zcheng\}@google.com}
  \country{}
}

\renewcommand{\shortauthors}{Ding, Kaize, et al.}
\begin{abstract}

Learning expressive representations for high-dimensional yet sparse features has been a longstanding problem in information retrieval. Though recent deep learning methods can partially solve the problem, they often fail to handle the numerous sparse features, particularly those tail feature values with infrequent occurrences in the training data. Worse still, existing methods cannot explicitly leverage the correlations among different instances to help further improve the representation learning on sparse features since such relational prior knowledge is not provided. To address these challenges, in this paper, we tackle the problem of representation learning on feature-sparse data from a graph learning perspective. Specifically, we propose to model the sparse features of different instances using hypergraphs where each node represents a data instance and each hyperedge denotes a distinct feature value. By passing messages on the constructed hypergraphs based on our Hypergraph Transformer (HyperFormer), the learned feature representations capture not only the correlations among different instances but also the correlations among features. Our experiments demonstrate that the proposed approach can effectively improve feature representation learning on sparse features.

\end{abstract}







\maketitle


\section{Introduction}


As one prevailing line of research for dealing with sparse features, researchers try to model the cross features in either the raw feature level \cite{rendle2010factorization} or the embedding level \cite{wang2021dcn} to improve the representation expressiveness. Despite their success, existing methods still have a major bottleneck of capturing the following relational information within data: (1)
\textit{Instance Correlations}. Most of the existing efforts assume training instances are independently and identically distributed (i.i.d.), while different instances in the data may share correlated behavior/feature patterns~\cite{zheng2022hien}. Since the features of an instance could be extremely sparse, leveraging the knowledge from other instances is highly beneficial to improve the representation quality of sparse features in a collective way. Yet, how to model the correlations between different data instances without prior knowledge remains unexplored in this field; (2) \textit{Feature Correlations}. In general, different feature values are usually correlated (e.g., two different feature values are commonly shared by many instances), which should not be neglected for learning expressive feature representations. 
In real-world systems, as the data scale grows continuously in a power-law distribution, a large number of features appear very few times in the training set, also known as tail features. As a result, existing methods such as feature interaction learning approaches that rely on feature co-occurrence will lose their efficacy on tail features due to their rarity in the data~\cite{guo2021dual}. How to better capture feature correlations is a key to improve the representation learning of sparse features. Based on those two limitations, one natural research question to ask is that – \textit{given the input data with high-dimensional and sparse features, is there a natural way to explicitly model both instance correlations and feature correlations simultaneously}?



To answer this question, we go beyond the existing learning paradigm and study relational representation learning on data with sparse features from a graph learning perspective. Due to the fact that the relationships among instances are naturally high-order rather than pair-wise, e.g., a group of users sharing the same feature "location", instead of using simple graph, we adopt hypergraph~\cite{zhou2006learning} to model the high-order correlations among data instances. Specifically, we take feature values as the proxy (i.e., hyperedges) to connect different data instances (i.e., nodes) with the same feature values. In order to facilitate learning representations on each hypergraph constructed from the feature-sparse data, we further develop a plug-and-play model -- Hypergraph Transformer (HyperFormer), which serves as an embedding module and is compatible to be trained together with arbitrary prediction model for task-agnostic sparse predictive analytics. HyperFormer iteratively aggregates the information from hyperedges to nodes and vice versa, which allows multi-hop message-passing on the constructed hypergraphs and captures the instance correlations as well as feature correlations simultaneously. The resulted feature representations can improve the predictive power of different models on feature-sparse data. Our experiments on (i) CTR prediction and (ii) top-K item recommendation tasks demonstrate that HyperFormer is generalizable across different tasks, and further enhances state-of-the-art approaches of representation learning for sparse features.

\section{Related Work}
\begin{figure}
\includegraphics[width=0.47\textwidth]{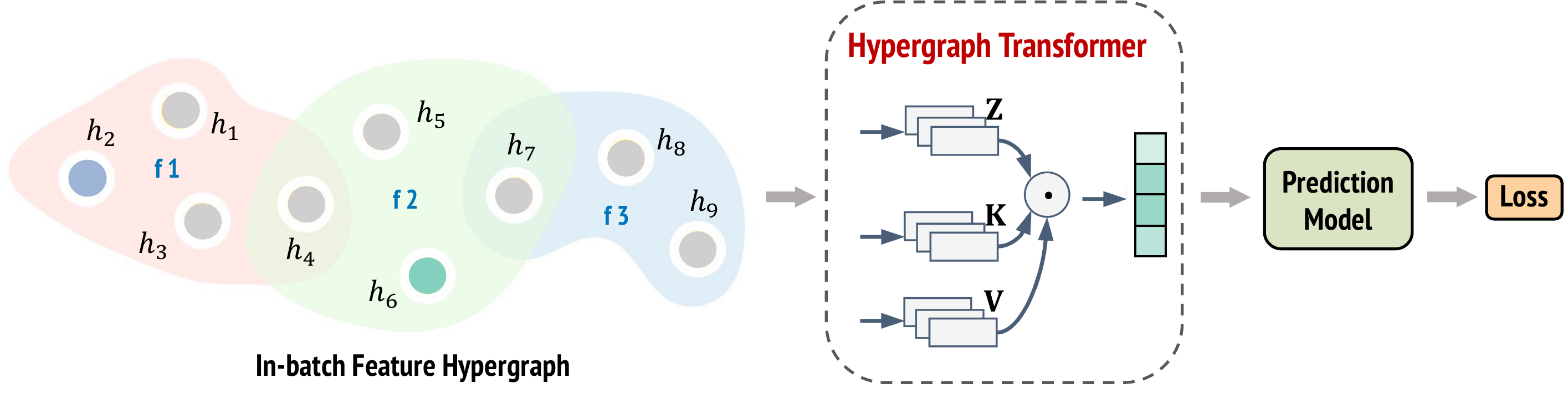}
\caption{Illustration of the proposed HyperFormer.}
\label{fig:hypergraph}
\end{figure}

\noindent\textbf{Learning with Sparse Features} has been a classic yet challenging problem in information retrieval and recommender system. 
A prevailing line of research tries to model the cross features in either the raw feature level or the embedding level. Compared to conventional approaches
~\cite{rendle2010factorization,juan2016field,pan2018field}, deep learning models have shown their superiority for handling high-dimensional sparse features~\cite{cheng2016wide,shan2016deep,wang2021dcn,ding2019feature}. Methods such as Wide\&Deep~\cite{cheng2016wide}, Deep Crossing~\cite{shan2016deep}, PNN~\cite{qu2016product}, DCN~\cite{wang2017deep}, AutoInt~\cite{song2019autoint}, Fi-GNN~ \cite{li2019fi}, DCN-v2~\cite{wang2021dcn} have been proposed to automatically model the cross features. However, existing methods are not able to explicitly capture the correlations between instances
and are also ineffective to handle the tail features that appear rarely in the data. 






\noindent\textbf{Graph Neural Networks (GNNs)} generally follow the neighborhood aggregation scheme~\cite{kipf2016semi,hamilton2017inductive,velickovic2017graph,ding2022data}, learning the latent node representation via message passing among local or high-order neighbors in the graph~\cite{klicpera2019predict,ding2022meta,ding2023structural}. 
More recently, GNNs have been actively explored to improve the performance of CTR prediction~\cite{li2019fi,li2021graphfm,guo2021dual,zheng2022hien} by capturing the interactions between features. Our approach leverages the idea of hypergraph~\cite{zhou2006learning,feng2019hypergraph,wang2020next,ding2020more,wang2021session,jeong2022nothing} and we build \textbf{HyperFormer} to perform representation learning on sparse features. In addition, our work focuses on developing a new embedding module rather than learning the cross-features, and our plug-and-play model is compatible to be trained with any feature interaction learning methods for making final predictions.








\section{Methodology}
\noindent\textbf{Problem Definition.}
For the sake of simplicity, we only consider sparse features and use multi-hot representation for sparse features, where each sparse feature is also called a field. The input of our problem is a high-dimensional sparse vector $\mathbf{x} \in \mathrm{R}^N$ of multi-hot representation, where $N$ is the total number of sparse feature values. Also, $x_i = 0$ means the $i$-th feature value does not exist in the instance and $x_i = 1$ means otherwise. The objective is to learn a low-dimensional embedding vector $\mathbf{e} \in \mathrm{R}^d$ that represents the raw input features in the latent space. 

Existing works apply an embedding layer to project the input features into a low dimensional feature vector, which is commonly implemented by looking up from an embedding table $\mathbf{F} = [\mathbf{f}_1, \mathbf{f}_2,..., \mathbf{f}_N] \in \mathbb{R}^{N \times d}$ and concatenating the retrieved embeddings into a dense real-value vector. Correspondingly, $\mathbf{f}_k$ can be regarded as the dense representation for feature $x_k$. In this paper, we argue that existing methods cannot explicitly consider the correlations between instances and the correlations between features, leading to the feature representations less expressive.

\subsection{Feature Hypergraph}
In this paper, we propose to alleviate the feature sparsity issue through relational representation learning. 
Since each specific feature can appear in multiple data instances, it can be naturally utilized as a bridge to capture instance correlations as well as feature correlations. For example, a group of users sharing the same feature values for ``location'' or ``age''. Such correlations among instances are inherently high-order rather than pair-wise, thus we propose to build \textit{feature hypergraph} to model the input data and try to enable message-passing on it to capture desired relational information. Specifically, we define the feature hypergraph as follows:
\begin{definition}
\textbf{Feature Hypergraph}: A feature hypergraph is defined as a graph $ G = (\mathcal{V}, \mathcal{E})$, where $\mathcal{V} = \{v_1, \dots, v_n\}$ represents the set of nodes in the graph, and $\mathcal{E} = \{e_1, \dots, e_m \}$ represents the set of hyperedges. Specifically, each node represents a data instance and each hyperedge represents a unique feature value. Correspondingly, for any hyperedge $e$, it can connect arbitrary number of nodes/instances (i.e., $\sigma(e) \geq 1$).
\end{definition}



\noindent\textbf{Scalability Extension.} Considering the fact that the scale of training data could be extremely large in practice, it is almost impossible to build a single feature hypergraph to handle all the data instances. To counter this issue, we propose to construct in-batch hypergraph based on data instances in the batch to further support mini-batch training. In Figure \ref{fig:hypergraph}, we illustrate the steps for constructing the in-batch hypergraphs. For each batch, we randomly sample a batch of instances and update the hypergraph structure based on the data samples in the batch. From our experiments, the in-batch feature hypergraph is also effective for capturing the desired data dependencies and achieves satisfying performance improvements.

\subsection{Hypergraph Transformer} 
To support representation learning on the constructed feature hypergraphs, we further propose a new model Hypergraph Transformer (HyperFormer) in this paper, which adopts the Transformer-like architecture \cite{vaswani2017attention} to exploit the hypergraph structure to encode both the instance correlations and feature correlations. Apart from conventional GNN models, each layer in HyperFormer learns the representations with two different hypergraph-guided message-passing functions, capturing high-order instance correlations and feature correlations simultaneously. Formally, a Transformer layer can be defined as:
\begin{equation}
\begin{aligned}
      \mathbf{H}^l &= \textsc{TF}_{edge}\Big( \mathbf{Q}_{edge}^l =  \mathbf{H}^{l-1}, \mathbf{K}_{edge}^l = \mathbf{F}^{l-1}, \mathbf{V}_{edge}^l = \mathbf{F}^{l-1}\Big),\\
    \mathbf{F}^{l}  &= \textsc{TF}_{node}\Big(\mathbf{Q}_{node}^l = \mathbf{F}^{l-1}, \mathbf{K}_{node}^l = \mathbf{H}^{l}, \mathbf{V}_{node}^l = \mathbf{H}^{l}\Big)
    \label{eq:graphSage2},
\end{aligned}
\end{equation}
where $\textsc{TF}\Big(\mathbf{Q}, \mathbf{K}, \mathbf{V}\Big) = \text{FFN}\Big[\text{softmax}(\frac{\mathbf{Q}\mathbf{K}^\mathrm{T}}{\sqrt{d}})\mathbf{V}\Big]$  denotes the Transformer-like attention mechanism. In essence, $\textsc{TF}_{edge}$ is a message-passing function that aggregates information from hyperedges to nodes and  $\textsc{TF}_{node}$ is another message-passing function that aggregates information from nodes to hyperedges. Specifically, we first look up from the feature embedding table $\mathbf{F}$ to initialize hyperedge representations and the initial node representation of each instance is computed by concatenating all its feature representations. Without loss of generality, we describe the two message-passing functions in a single HyperFormer layer $l$ as follows:

\noindent\textbf{Feature-to-Instance Message-Passing.} 
With all the hyperedges representations $\{\mathbf{f}_j^{l-1} \arrowvert \forall e_j \in \mathcal{E}_i\}$, we first apply an feature-to-instance (edge-to-node) message-passing to learn the next-layer representation  $\mathbf{h}_i^l$ of node $v_i$. Specifically, we set the node representation from the last HyperFormer layer $l-1$ as the query. The representations of the connected hyperedges can be projected into keys and values. Formally, the similarity between the query and key can be calculated as:
\begin{equation}
\begin{aligned}
    \alpha_{ij} &= \frac{\exp((\mathbf{h}_i^{l-1}\mathbf{W}^Q_{edge})^{\mathrm{T}} \mathbf{k}_{j})}
     {\sum_{e_p \in \mathcal{E}_i}\exp((\mathbf{h}_i^{l-1}\mathbf{W}^Q_{edge})^{\mathrm{T}} \mathbf{k}_{p})},  \quad\mathbf{k}_{p} = \mathbf{f}_{p}^{l-1}\mathbf{W}^K_{edge},
\end{aligned}
\label{equ:att_node2edge}
\end{equation}
in which $\mathbf{W}^K_{edge}$ is the projection matrix for the key of the feature-to-instance transformer. Then the next layer node representation can be computed as:
\begin{equation}
    \mathbf{h}_i^l = \sigma\bigg( \sum_{e_j \in \mathcal{E}_i} \alpha_{ij} \mathbf{f}_{j}^{l-1}\mathbf{W}^V_{edge} \bigg),
\end{equation}
where $\sigma$ is the non-linearity such as ReLU and $\mathbf{W}^V_{edge}$ is a trainable projection matrix for the value.



\noindent\textbf{Instance-to-Feature Message-Passing.} 
With all the updated node representations, we again apply an instance-to-feature (node-to-edge) message-passing based on the Transformer layer to learn the next-layer representation of hyperedge $e_j$. Similarly, this process can be formally expressed as:
\begin{equation}
    \mathbf{f}_{j}^{l} = \sigma\bigg( \sum_{v_k \in \mathcal{V}_j} \beta_{jk} \mathbf{h}_k^l\mathbf{W}^V_{node} \bigg), 
\end{equation}
where $\mathbf{f}_{j}^{l}$ is the output representation of hyperedge $e_j$ and $\mathbf{W}^V_{node}$ is the projection matrix. $\beta_{jk}$ denotes the attention score of hyperedge $e_j$ on node $v_k$, which can be computed by: 
\begin{equation}
\begin{aligned}
    \beta_{jk} &= \frac{\exp((\mathbf{f}_{j}^{l-1}\mathbf{W}^Q_{node})^{\mathrm{T}} \mathbf{k}_{k})}
     {\sum_{v_p \in \mathcal{V}_j}\exp(\mathbf{f}_{j}^{l-1}\mathbf{W}^Q_{node})^{\mathrm{T}} \mathbf{k}_{p})},   \quad \mathbf{k}_{p} = \mathbf{h}_p^l\mathbf{W}^K_{node}, 
\end{aligned}
\label{equ:att_edge2node}
\end{equation}
where $\mathbf{W}^Q_{node}$ and $\mathbf{W}^K_{node}$ are the projection matrices for the query and key of the instance-to-feature message-passing. By stacking multiple HyperFormer layers, we are able to capture high-order instance correlations and feature correlations. The feature representations learned from the last HyperFormer $\mathbf{F}^L$ can be directly plugged into any model architecture as the feature embedding layer and improve the prediction performance on the downstream tasks.


\section{Experiments}
To evaluate the effectiveness of the proposed approach, we conduct our experiments on two real-world tasks that often suffer from the feature sparsity issue: (i) click-through-rate (CTR) prediction and (ii) top-k item recommendation. 
\subsection{Task1: CTR Prediction}
Click-through-rate (CTR) prediction is a task that predicts how likely a user is going to click an advertisement. Typically, an instance sample is represented by high-dimensional and sparse features, such as user profile, ad attributes, and contextual features such as time, platform, and geographic location. We first try to evaluate the effectiveness of HyperFormer for CTR prediction.


\noindent\textbf{Datasets.} For the task of CTR prediction, we adopt two public real-world benchmark datasets in which the features are extremely sparse and the statistics for those datasets can be found in Table \ref{table:ctr_data}. To adopt 
the \textbf{MovieLens-1M} dataset for CTR prediction, we follow \cite{wang2021dcn,song2019autoint} to transform the original user ratings into binary values. 
The dataset is divided into 8:1:1 for training, validation, and testing, respectively.
The \textbf{Criteo} dataset is widely adopted for CTR prediction that includes 45 million users’ ad clicks on display ads over a 7-day period. As in \cite{wang2021dcn,song2019autoint}, we use the data from the first 6 days for training and randomly split the data from the last day into validation and test sets.

\begin{table}[t]
\centering
\caption{Dataset Statistics.}
\scalebox{0.8}{



\begin{tabular}{c|c|ccc}
\toprule
\multirow{3}{*}{\textbf{CTR Prediction}} & Dataset & \#Sample & \#Field & \#Feature  \\ \cline{2-5} 
 & MovieLen-1M & 0.94M & 7 & 3,529  \\
 & Criteo & 45.84M & 39 & 998,960  \\ \hline
 \midrule
\multirow{3}{*}{\textbf{Item Recommendation}} & Dataset & \#User & \#Item & \#Features \\ \cline{2-5} 
 & Amazon-Movie & 19,873 & 10,176 &  8,504 \\
 & Bookcrossing & 48,999  & 193,765  & 5,100 \\ 
  \bottomrule
\end{tabular}
}
\label{tab:classification_data}
\label{table:ctr_data}
\end{table}







\noindent\textbf{Baselines.} We include the following baselines methods for CTR prediction: Logistic Regression (\textbf{LR}) and Factorization Machine (\textbf{FM}) \cite{rendle2010factorization}. Different neural extensions of FM, including Neural Factorization Machine (\textbf{NFM}) \cite{he2017neural}, \textbf{xDeepFM} \cite{lian2018xdeepfm}, and \textbf{HoFM} \cite{blondel2016higher}. \textbf{AutoInt} \cite{song2019autoint} is designed to automatically learn the feature interaction with self-attention. \textbf{DCN-v2} \cite{wang2021dcn} is an improved deep \& cross network that models the explicit and bounded-degree feature interactions. To show the flexibility and effectiveness of HyperFormer, we integrate it into two representative baselines AutoInt and DCN-v2 then report their performance in the experiments.


\noindent\textbf{General Comparison.} We evaluate the  performance of different methods based on two widely-used metrics for CTR prediction: AUC and LogLoss in Table \ref{table:ctr}. FM is able to model the second-order feature interaction and thus outperforms LR, which can only learn from raw feature input. With the power of deep neural networks, xDeepFM and NFM can improve the performance of FM in both datasets by incorporating non-linear transformations and interactions among features. AutoInt further improves on NFM by adaptively modeling feature interactions using an attention mechanism. DCN-v2 is also shown to be an effective approach for CTR. More importantly, our experimental results demonstrate the effectiveness of HyperFormer, as it improves the performance of the two representative CTR models for both AUC and LogLoss.

\begin{figure}[!h]
    \centering
    \subfigure 
    {
    \includegraphics[width=0.23\textwidth]{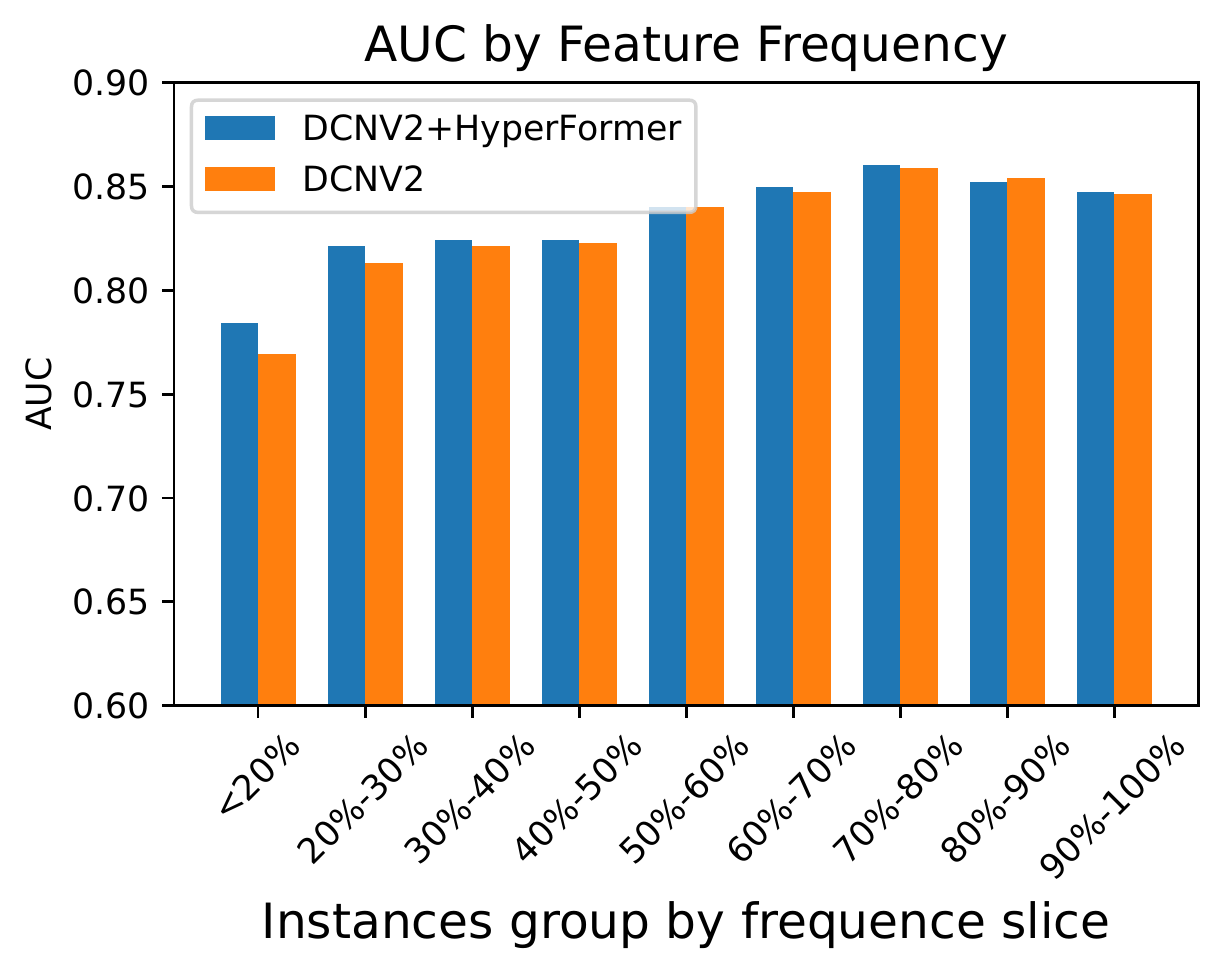}
    }
    \hspace{-0.35cm}
    \subfigure
    {
    \includegraphics[width=0.23\textwidth]{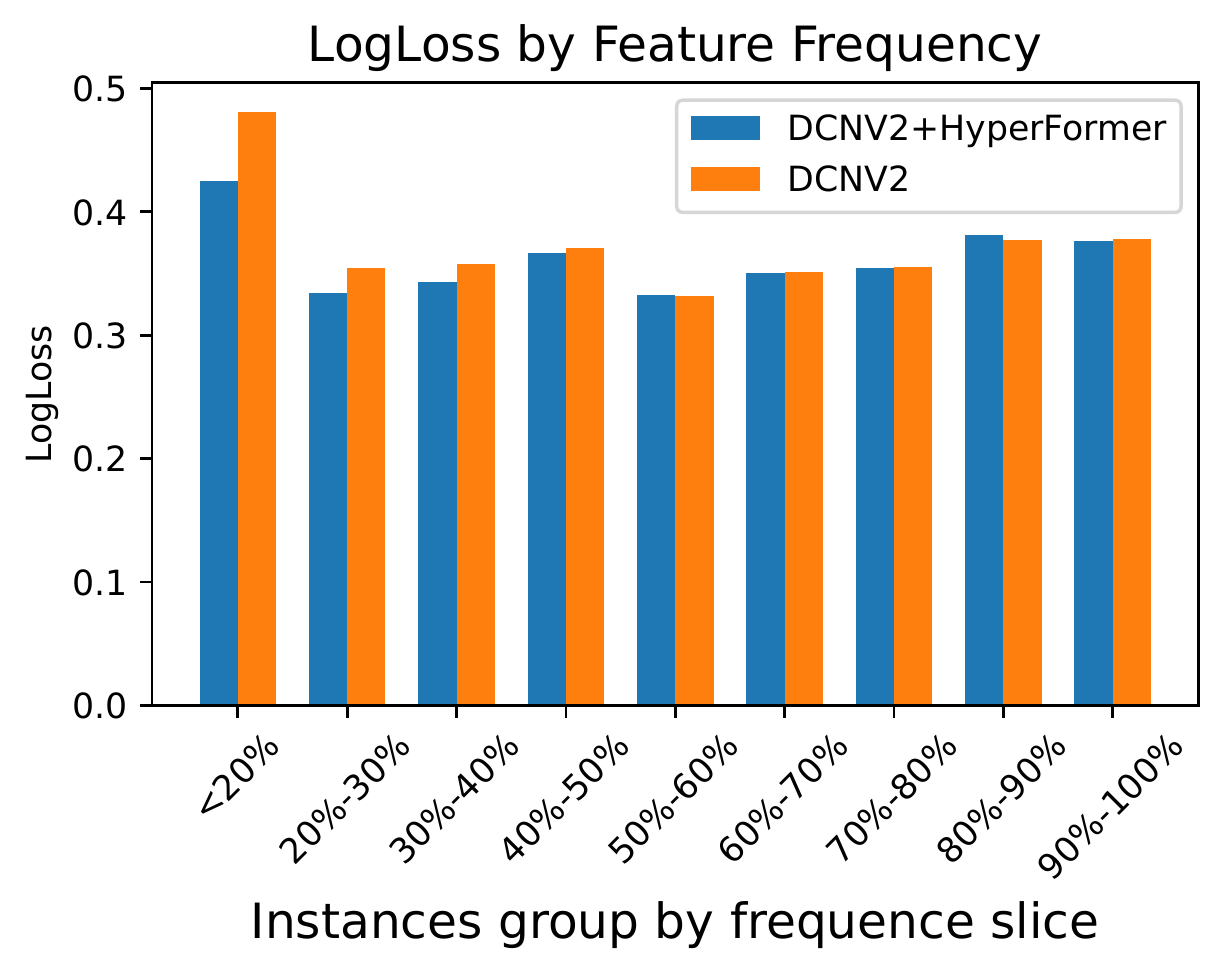}
    }
    \caption{CTR performance across different instance groups.}%
    \label{fig:tail_features}
\end{figure}

\noindent\textbf{Further Analysis on Tail Features.}  Feature values usually follow a power-law distribution and those tail features only appear a few times among all the data examples. Without enough learning signals, it is hard for the low-frequency features to obtain informative embeddings, resulting in low CTR prediction accuracy for data samples that contain those low-frequency features. Our HyperFormer is proposed to address this issue by modeling the correlations between features through hypergraph message passing. 
To examine whether HyperFormer achieves this goal, we first sort and slice all the feature values in MovieLens-1M by frequency. Then we retrieve the test inputs that contain each set of feature values for evaluation. We compare the CTR performance of DCN-v2 with and without HyperFormer in Figure \ref{fig:tail_features}.  Our results show that HyperFormer enables DCN-v2 to achieve better performance on samples with low-frequency features, as measured by both LogLoss and AUC. This demonstrates that HyperFormer is effective in enhancing the quality of feature embeddings for tail features, leading to more accurate CTR predictions for items with tail features.

\begin{table}[t!]
\centering
\caption{AUC and Logloss on CTR prediction.}
\scalebox{0.875}{
\begin{tabular}{@{}lcccccc@{}}
\toprule

\rule{0pt}{10pt} \textbf{Method} & \multicolumn{2}{c}{Movielens-1M}  & & \multicolumn{2}{c}{Criteo} \\

&  AUC  & LogLoss & & AUC  & LogLoss

\\ \midrule



LR   &$0.7716$& $0.4424$ && $0.7820$ & $0.4695$   \\
FM   &$0.8252$& $0.3998$ && $0.7836$ & $0.4700$  \\
NFM &$0.8357$& $0.3883$&& $0.7957$ & $0.4562$\\
xDeepFM   &$0.8286$& $0.4108$ && $0.8009$ & $0.4517$   \\
HoFM  & $0.8304$ & $0.4013$ && $0.8004$  & $0.4508$  \\
AutoInt  & $0.8456$ & $0.3797$ && $0.8061$  & $0.4455$  \\
DCN-v2    & $0.8402$   & $0.3811$ && $0.8045$ & $0.4462$\\

\midrule

 \textbf{AutoInt+HyperFormer}  & $\mathbf{0.8462}$   & $\mathbf{0.3770}$ && $\mathbf{0.8072}$ &  $\mathbf{0.4444}$\\
  \textbf{DCN-v2+HyperFormer}  & $\mathbf{0.8471}$   & $\mathbf{0.3755}$ && $\mathbf{0.8061}$  & $\mathbf{0.4453}$\\

\bottomrule
\end{tabular}}
\label{table:ctr}
\end{table}




\subsection{Task2: Top-K Item Recommendation}
In many real-world recommendation systems, the goal is to retrieve the Top-K relevant items for a given user. Indeed, the input of these top-k recommendation frameworks also suffers from the same issue as the CTR task in that the user and item features are also extremely sparse and follow the long-tailed distribution. To further demonstrate the generalizability of HyperFormer, we evaluate its performance on the relational representation learning for recommendation systems.

\noindent\textbf{Datasets.} In the experiment, we adopt two benchmark datasets for evaluation. \textbf{Amazon-Movie} consists of product reviews and metadata for the "Movie" category spanning from May 1996 to July 2014 \cite{mcauley2015image}. \textbf{Bookcrossing} \cite{ziegler2005improving} collects the user-item ratings within the community, including both user demographic and age as well as item features such as Title, Author, Year, and Publisher. After removing the inactive users and items, we obtain the final datasets as summarized in Table \ref{table:ctr_data}. We randomly sample 70\% data for model training, 10\% for validation and 20\% for testing.

\noindent\textbf{Baselines.} Due to its high efficiency and flexibility, \textbf{Two-tower} model with separate user and item towers is widely adopted as the fundamental learning architecture for large-scale top-k item retrieval \cite{yi2019sampling,yao2021self,wang2021cross,yu2021dual,zhang2021model}. Specifically, the high-dimensional user and item features are input to the corresponding towers, and the preference scores are typically computed by a dot-product between user and item embeddings encoded by the corresponding towers. To solve the feature imbalance issue, \textbf{DAT} \cite{yu2021dual} was proposed to extend each tower with an extra learnable vector to memorize the cross-tower information. Recently, Yao et.al proposed \textbf{SSL} \cite{yao2021self} to leverage latent feature correlations in a two-tower model by augmenting the data and incorporating an auxiliary self-supervised learning task.

\begin{figure}[!h]
    \centering
    \subfigure
    {
    \includegraphics[width=0.225\textwidth]{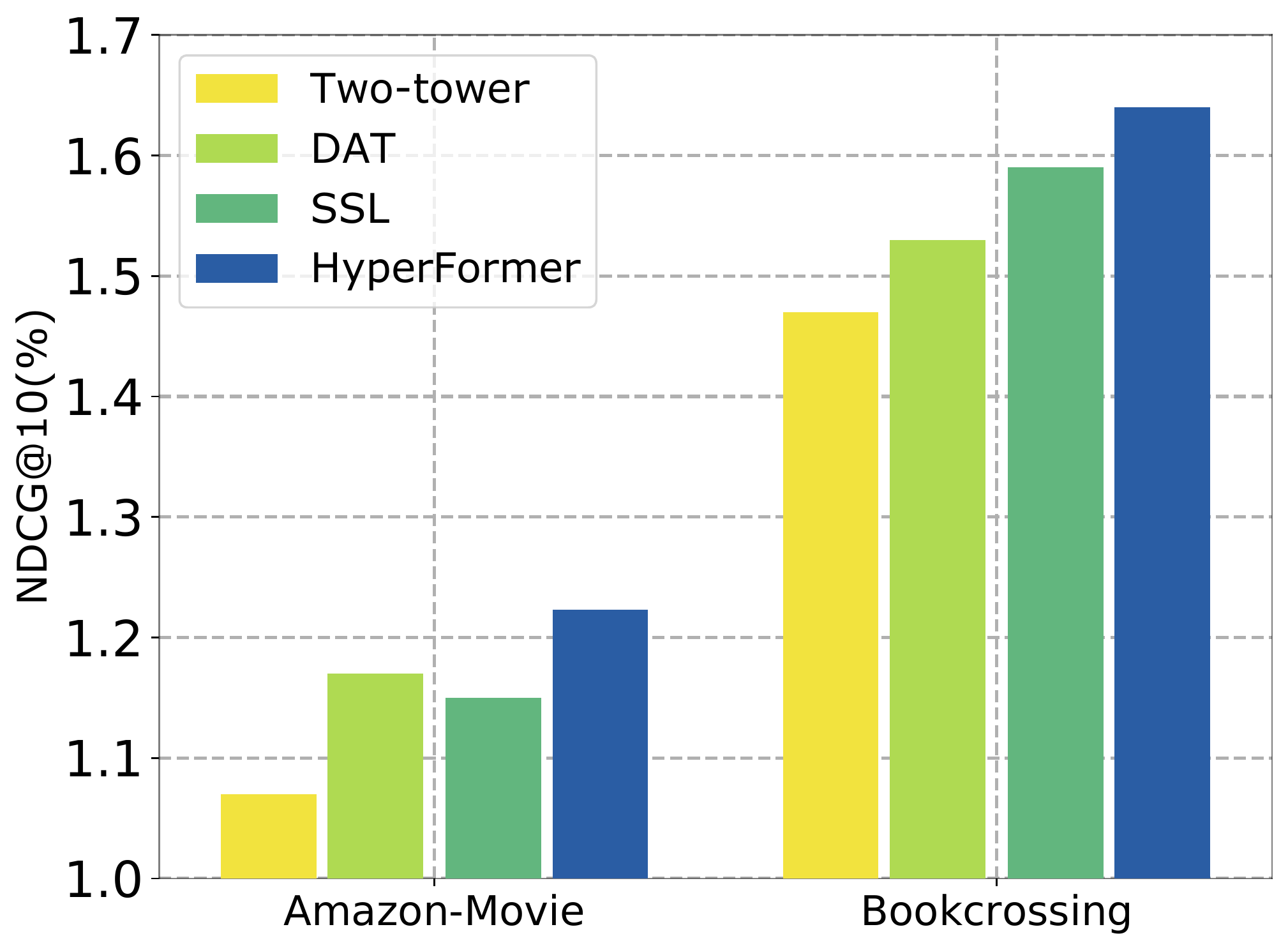}
    }
    \hspace{-0.1cm}
    \subfigure
    {
    \includegraphics[width=0.225\textwidth]{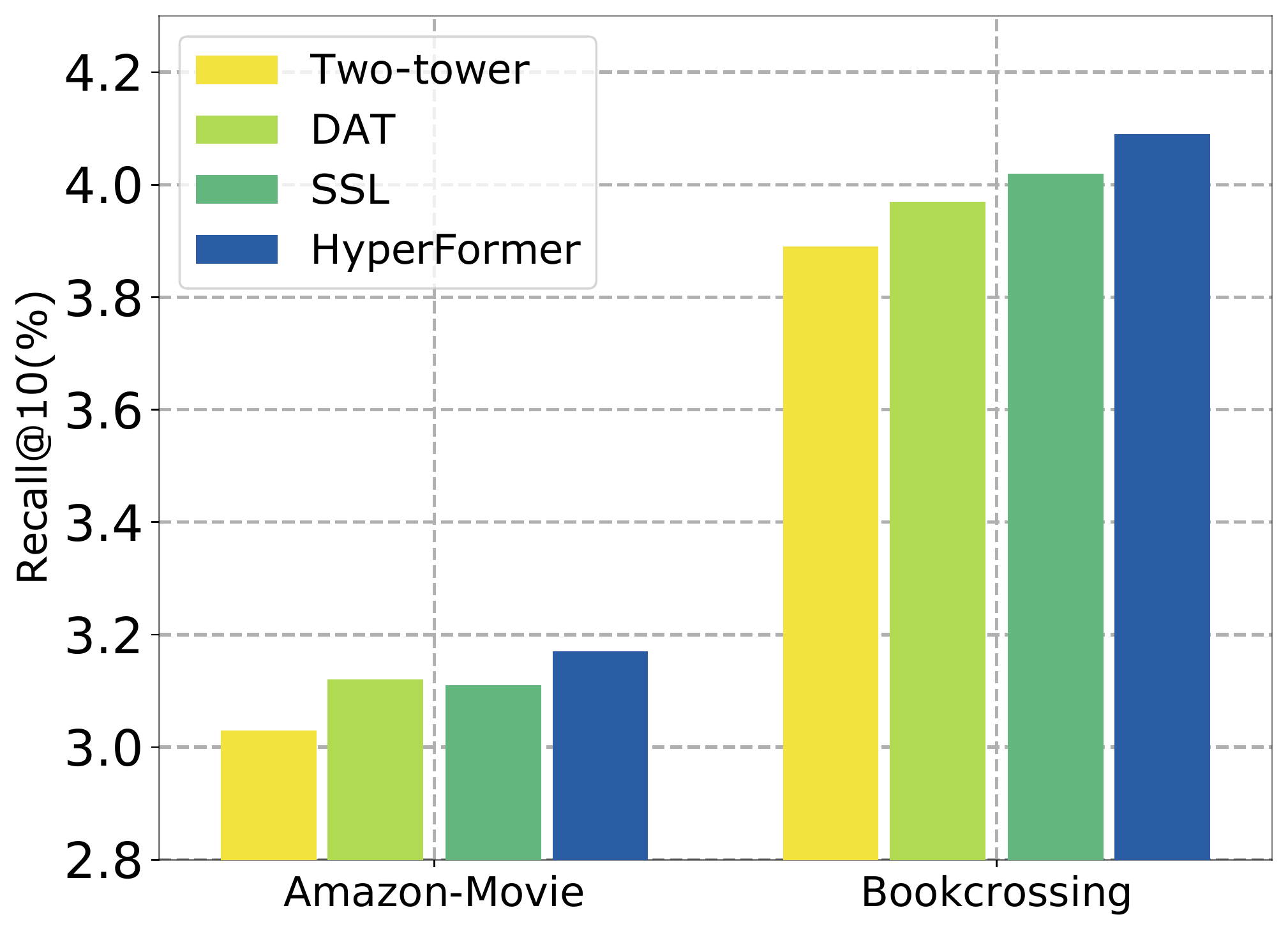}
    }
    \caption{Top-k Recommendation performance comparison.}%
    \label{fig:topk}
    \label{fig:batch_size}
\end{figure} 

\noindent\textbf{General Comparison.} To evaluate HyperFormer on top-K item recommendation, we plug it into a two-tower recommendation model and compare it with the baseline methods above that were proposed to address the feature sparsity issue. We use NDCG@10 and Recall@10 as evaluation metrics and summarize the results for both Amazon-Movie and Bookcrossing in Figure \ref{fig:topk}. By introducing the category alignment loss and an extended vector capturing the cross-tower information, DAT can significantly improve two-tower in top-k recommendation for both datasets. We find that SSL significantly outperforms DAT in Bookcrossing but falls behind DAT in Amazon-Movie. However, the proposed HyperFormer consistently outperforms all other methods in both datasets, showcasing its effectiveness in feature representation learning. The advantage across both CTR prediction and top-k recommendation highlights the generalizability of HyperFormer and its potential to address feature sparsity in various real-world tasks.

\section{Conclusion}
In this paper, we focus on the problem of representation learning on high-dimensional sparse features. We propose to build feature hypergraphs to model the instance correlations and feature correlations explicitly. The proposed Hypergraph Transformer further enables message-passing on the constructed feature hypergraphs, resulting in more informative feature representations that encode instance correlations and feature correlations within the data. The evaluation of different methods on click-through-rate prediction and item recommendation demonstrate the effectiveness of our approach in capturing the relational information within data for learning informative feature representations.

\section*{Acknowledgments}

This material is based upon work supported by, or in part by, the
NSF grant 2229461.

\balance
\bibliographystyle{plain}
\bibliography{acmart}

\end{document}